\documentclass[sigconf]{acmart}
\usepackage{subfig}
\usepackage{graphicx}
\usepackage{color}
\usepackage{amsmath}
\usepackage{multirow}
\usepackage{booktabs}
\usepackage{url}
\usepackage{amsmath}
\usepackage{adjustbox}
\usepackage{subfloat}
\usepackage[english]{babel}
\usepackage{comment}
\usepackage{enumitem}
\usepackage{tikz}

\AtBeginDocument{%
  \providecommand\BibTeX{{%
    \normalfont B\kern-0.5em{\scshape i\kern-0.25em b}\kern-0.8em\TeX}}}

\acmYear{2024}\copyrightyear{2024}
\acmConference[MMVE '24]{16th International Workshop on Immersive Mixed and Virtual Environment Systems}{April 15--18, 2024}{Bari, Italy}
\acmBooktitle{16th International Workshop on Immersive Mixed and Virtual Environment Systems (MMVE '24), April 15--18, 2024, Bari, Italy}
\acmDOI{10.1145/3652212.3652222}
\acmISBN{979-8-4007-0618-9/24/04}

\setcopyright{acmlicensed}
\copyrightyear{2024}
\acmYear{2024}
\acmDOI{10.1145/3652212.3652222}

\begin{document}

%%
%% The "title" command has an optional parameter,
%% allowing the author to define a "short title" to be used in page headers.
\title{Influence of Gameplay Duration, Hand Tracking, and Controller Based Control Methods on UX in VR}
%%\title{Assessing the Impact of Gameplay Duration and Hand Tracking vs. Controller Methods on UX in VR}
%% \title{Impact of Gameplay Duration and Control Methods on User Experience in Virtual Reality}

%%
%% The "author" command and its associated commands are used to define
%% the authors and their affiliations.
%% Of note is the shared affiliation of the first two authors, and the
%% "authornote" and "authornotemark" commands
%% used to denote shared contribution to the research.

\author{Tanja Kojić}
\affiliation{\institution{Quality and Usability Lab\\Technical University of Berlin} \country{Germany}}

\author{Maurizio Vergari}
\affiliation{\institution{Quality and Usability Lab\\Technical University of Berlin} \country{Germany}}

\author{Simon Knuth}
\affiliation{\institution{Quality and Usability Lab\\Technical University of Berlin} \country{Germany}}

\author{Maximilian Warsinke}
\affiliation{\institution{Quality and Usability Lab\\Technical University of Berlin} \country{Germany}}

\author{Sebastian Möller}
\affiliation{\institution{Quality and Usability Lab\\Technical University of Berlinand German Research Center for Artificial Intelligence (DFKI)} \country{Germany}}

\author{Jan-Niklas Voigt-Antons}
\affiliation{\institution{Immersive Reality Lab\\Hamm-Lippstadt University of Applied Sciences} \country{Germany}}

\makeatletter
\let\@authorsaddresses\@empty
\makeatother

\begin{comment}

\author{Tanja Kojić}
\affiliation{%
 \institution{Quality and Usability Lab, Technical University Berlin}
 \country{Germany}}

\author{Maurizio Vergari}
\affiliation{%
 \institution{Quality and Usability Lab, Technical University Berlin}
 \country{Germany}}

\author{Maximilian Warsinke}
\affiliation{%
 \institution{Quality and Usability Lab, Technical University Berlin}
 \country{Germany}}

\author{Sebastian Möller}
\affiliation{%
 \institution{Quality and Usability Lab, Technical University Berlin \& DFKI}
 \country{Germany}}
 
\end{comment}
 
%%
%% By default, the full list of authors will be used in the page
%% headers. Often, this list is too long, and will overlap
%% other information printed in the page headers. This command allows
%% the author to define a more concise list
%% of authors' names for this purpose.
%\renewcommand{\shortauthors}{Trovato and Tobin, et al.}

%%
%% The abstract is a short summary of the work to be presented in the
%% article.
\begin{abstract}
Inside-out tracking is growing popular in consumer VR, enhancing accessibility. It uses HMD camera data and neural networks for effective hand tracking. However, limited user experience studies have compared this method to traditional controllers, with no consensus on the optimal control technique. This paper investigates the impact of control methods and gaming duration on VR user experience, hypothesizing hand tracking might be preferred for short sessions and by users new to VR due to its simplicity. Through a lab study with twenty participants, evaluating presence, emotional response, UX quality, and flow, findings revealed control type and session length affect user experience without significant interaction. Controllers were generally superior, attributed to their reliability, and longer sessions increased presence and realism. %Surprisingly, those with more VR experience preferred hand tracking, challenging initial assumptions.
The study found that individuals with more VR experience were more inclined to recommend hand tracking to others, which contradicted predictions.
\end{abstract}

%%
%% The code below is generated by the tool at http://dl.acm.org/ccs.cfm.
%% Please copy and paste the code instead of the example below.
%%


%%
%% Keywords. The author(s) should pick words that accurately describe
%% the work being presented. Separate the keywords with commas.

%% A "teaser" image appears between the author and affiliation
%% information and the body of the document, and typically spans the
%% page.
\begin{comment}    
\begin{teaserfigure}
  \includegraphics[width=\textwidth]{sampleteaser}
  \caption{Seattle Mariners at Spring Training, 2010.}
  \Description{Enjoying the baseball game from the third-base
  seats. Ichiro Suzuki preparing to bat.}
  \label{fig:teaser}
\end{teaserfigure}
\end{comment}

%\received{20 February 2007}
%\received[revised]{12 March 2009}
%\received[accepted]{5 June 2009}

%%
%% This command processes the author and affiliation and title
%% information and builds the first part of the formatted document.
\maketitle

\section{Introduction}
Virtual Reality (VR) headsets are rapidly gaining popularity, with sales expected to triple in three years by 2027~\cite{statista}. The development of standalone head-mounted displays (HMDs) with inside-out tracking has surely contributed to their success. This device type opens up the platform to a wider audience by eliminating the need for advanced technical abilities, high-performance PCs, and sensor setups.
VR headsets, like game consoles, are typically controlled using specialised handheld controllers.
These controllers allow for low-latency interaction with 3D material by tracking them in space. Immersion in virtual environments relies heavily on input, with more natural-looking input leading to higher levels of immersion~\cite{mcgloin2013video}. VR systems try to simulate real-world interactions as precisely as feasible. The next step for input is to eliminate controllers and use only hand tracking. 
Modern headsets include integrated cameras for inside-out tracking, which can be used to track hands and fingers with great precision~\cite{han2020megatrack}. The Meta Quest platform demonstrates this capability.
Using hand tracking instead of a controller can lower the barrier of entry for those unfamiliar with VR, eliminating the need for button mappings.
Previous evaluations of these control systems in user experience (UX) have shown inconsistent results~\cite{khundam2021comparative, hameed2021evaluating, voigt2020influence}, suggesting another element may be at play. This paper aims to explore the impact of gameplay time in relationship with the control method, as well as assessing users' willingness to interact with technology to see if those who are more open to new systems~\cite{franke2019personal} are more likely to be convinced by hand tracking, given the limitations of current methods.

\subsection{Gameplay Duration and Technology Affinity}

In the field of user study design, it is usual practice to keep the duration of the experience uniform throughout the experiment. However, there have been cases where researchers deviated from this pattern, undertaking studies that investigated the impact of different experience durations on user satisfaction and engagement. For example, in one study, participants engaged with a VR game for 2 and 5 minutes, providing insight into the possible implications of time on user experience. Intriguingly, the data revealed a potential link between longer durations and increased flow experiences, though it should be noted that this study did not provide thorough insights into other critical aspects of user experience \cite{volante2018time}.

In order to explore more into the world of user adaptation to innovative technologies, it becomes clear that the process frequently necessitates a time and effort investment on the part of the user. This investment might vary greatly depending on things such as previous experiences and personal character features. Some users are naturally hesitant to face the obstacles of unknown technology, but others are eager to embrace and study new systems and functionalities in order to solve problems more efficiently. The Affinity for Technology Interaction (ATI) questionnaire accurately captures and models these two conflicting preferences \cite{franke2019personal}.
Interestingly, while the ATI questionnaire is a repeating component of user studies involving participants' interactions with technology, it is rarely used as an independent variable in research designs. %This opens up a gap for future research, since understanding the role of users' affinity for technology engagement in shaping their experiences could provide useful insights into the design and implementation of user-centered technologies in the future, including one for VR systems.

%\newpage

\subsection{Hand Tracking as Control Method} %
Inside-out tracking technology is widely used in consumer VR systems, with recent examples including the Meta Quest and Pico. This system uses data from a series of integrated cameras to detect complex hand motions in three-dimensional space in real time. Hand tracking's exceptional accuracy makes it a very practical way to navigate some of the different virtual environments.
However, despite many advantages, hand tracking technology is not without disadvantages. 

One prominent challenge is the complexity of hand-to-hand interactions and the tracking of unusual hand positions \cite{han2020megatrack}. The headset's camera system has a limited field of view, which is a major concern. While many hand activities occur directly in front of the user's face, because they are naturally focused on these actions, some tasks, particularly those that replicate natural movements, take place in the peripheral and lower fields of vision.
This offers an important challenge because gestures conducted in these locations may not be efficiently caught by as many cameras, resulting in a decrease in tracking accuracy. In rare situations, specific movements may fall totally outside the scope of the cameras, resulting in a complete loss of tracking functionality \cite{buckingham2021hand}.
As a result, while inside-out tracking technology has transformed VR interaction, overcoming the issues associated with field of view constraints and guaranteeing consistent tracking precision in all hand positions remains an important focus of research and development within the VR industry.

While hand tracking technology is not without its limitations, it presents an promising potential for enhancing immersion within virtual reality environments. One of its main advantages is its ability to improve the user experience by expanding the range of "natural sensorimotor contingencies for perception" offered by VR systems \cite{slater2018immersion}.
Modern VR controllers, while effective in many respects, are limited by their design. They can only track certain parts of the hands, such as individual fingers, and restrict the range of hand poses that users can perform while holding them. This constraint is especially apparent when considering scenarios involving complicated 3D manipulation activities, in which the VR system must collect and duplicate details of these manipulations \cite{riecke20183d}.
In these cases, hand tracking technology appears as a more appealing option than controllers, at least when physical feedback is not the major concern. Hand tracking's capacity to accurately simulate natural hand movements and gestures has the potential to provide users with a more intuitive and immersive virtual experience. %As the VR industry evolves, addressing hand tracking constraints and improving its capabilities may lead to more interesting and immersive VR interactions in the future.

%\newpage
Several studies have investigated the potential of hand tracking technologies in virtual reality (VR), resulting in a complex findings and suggestions. In one such study, which intended to determine the comparative usefulness and satisfaction levels of VR controllers and hand tracking within a medical training simulation, no significant differences were discovered \cite{khundam2021comparative}.
In contrast, another study investigated the effectiveness of these two control approaches in carrying out simple reach-pick-place tasks. Surprisingly, the results favoured controllers, both in terms of objective performance measures and participant subjective ratings \cite{hameed2021evaluating}. This finding highlights the complex character of the hand tracking vs. controller argument, implying that the choice between different control systems may be determined by the unique environment and job at hand.

Furthermore, a third study added another degree of complexity to this discussion by concluding that, while hand tracking technology resulted in a more positive overall user experience, it appeared to come at the expense of lower perceived dominance in contrast to controllers \cite{voigt2020influence}. This intriguing contradiction highlights the complex nature of the comparison between these two control modalities, implying that factors other than usability may influence the final preference for one over the other.
It is important to note that, while these studies provide useful insights, they only partially overlap in their judgements, and they do not give a clear consensus on how hand tracking technology compares to traditional controllers in the VR landscape. As a result, additional study and a thorough examination of the unique situations and user preferences will be needed to untangle the details of this ongoing research and reach more definitive findings.

\subsection{Objectives}
In terms of VR games, hand tracking and controllers both offer advantages and disadvantages. Previously mentioned studies have been conducted to determine how they affect the user experience, but few have examined how the handling mechanism and game length interact. 

%Hand tracking technology appears to be promising, but it has yet to gain widespread use in consumer electronic devices. It is also interesting to understand how users' ATI may influence their willingness to adopt hand tracking. A higher ATI score may indicate that someone is more eager to attempt this technology, even if there are some technical issues. This could affect its popularity and utilisation in virtual reality experiences.
While hand tracking technology has huge potential, it has yet to become widely used in consumer products. This raises interesting questions about the factors influencing its adoption. Specifically, it is important to explore the effect of users' ATI in shaping users willingness to embrace hand tracking. A higher ATI score may indicate an increased interest to investigate and experiment with this technology, regardless of its occasional technical difficulties. As a result, understanding how ATI interacts with the popularity and use of hand tracking in VR experiences aims to give insight into its future direction in the constantly evolving arena of VR technology.

Therefore two research questions have been created for this paper as: 

\begin{itemize}
    \item How do control method and gameplay duration influence user experience for virtual reality gaming, and is there an interaction effect between the two?
    \item Is hand tracking more popular with people who are less experienced with virtual reality, and does this preference vary over different gameplay durations?
\end{itemize}

\newpage
\section{Methodology}

\begin{figure*}[h!]
  \centering
   \subfloat[Cubism]{
     \includegraphics[]{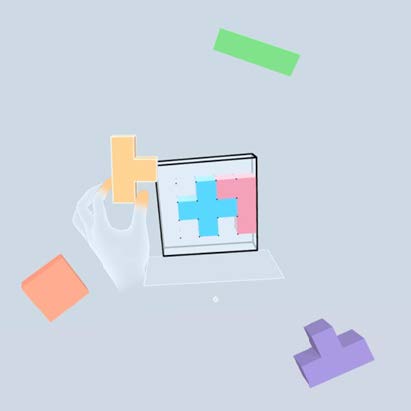}
    }
  \subfloat[Vacation Simulator]{
    \includegraphics[]{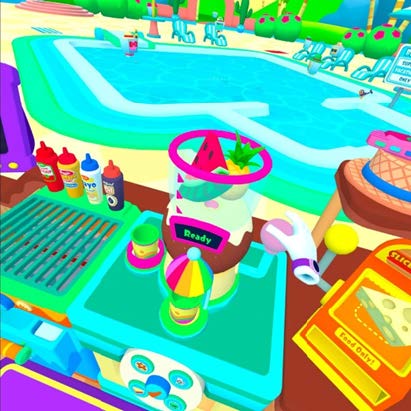}
    }
    \caption{Screenshots of games: a) A scene from Cubism depicting a partially solved puzzle with some shapes inserted, one in the player’s left hand and additional shapes floating in the play space, b) A scene from Vacation Simulator: Back to Job depicting the player pulling a lever on a stylized blender that’s filled with fruits while standing in a kitchen with a pool in the background.}
    \label{fig: games}
\end{figure*}

The study was mainly aimed to evaluate the user experience in VR and to collect additional data as needed to answer our research questions. To ensure consistency, we carried out the experiment in a controlled laboratory setting on the university campus. For the practical VR component of the research, we chose two different gaming durations and two separate control approaches. Each period related to a separate VR game.

The shorter duration lasted three minutes and included the puzzle game Cubism. We restart the game for each player, beginning with the first levels. Each level introduced a new three-dimensional geometric form that players had to "assemble" using a predetermined set of smaller pieces. The solution options were restricted, but both the goal form and the smaller shapes could be freely rotated and shifted, expanding the number of potential locations. All of the forms floated inside the available area, and players were free to shift their viewpoint. Interactions focused mostly on grabbing moving, rotating, and releasing shapes.
The longer duration lasted nine minutes and included the interactive narrative game Vacation Simulator. Like the shorter game, we reset it for each player, beginning with a special lesson.
Participants played the game's "Back to Job" option, which provided an infinite simulation of a receptionist's job at a holiday resort. The space was restricted to a main workstation and a kitchen area. Players were required to complete the resort's visitors' basic demands by discovering and interacting with the appropriate items in their surroundings. The game has an episodic format, with each episode comprising one or more main tasks. A virtual screen showed visuals of what participants should be looking for and what actions they needed to do with the items. Interactions mostly consisted on grabbing, moving, rotating, and releasing items, with rare interactions with virtual buttons.

The average session time for VR headsets, excluding those that rely on mobile phones, is about 46 minutes \cite{statistatime}. To avoid possible VR-induced symptoms and consequences, it is recommended that session lengths be limited to 55 to 70 minutes while conducting user research in VR \cite{kourtesis2019validation}. Given that many participants may be experiencing VR for the first time, it is best to aim for somewhat shorter periods to avoid problems.
Given the necessity for participants to complete surveys and follow instructions, the overall study time was set at around 60 minutes. A bit less than half of this time was spent within the VR headset, comfortably falling below of the limit.

The study's control techniques included two options: controllers and hand tracking.
This research used the Meta Quest 2 (previously known as Oculus Quest 2), a commercial VR headset with six degrees of freedom. This independent headset removes the need for any attached connections, giving users a great deal of mobility. With a per-eye resolution of 1832x1920 and built-in audio output speakers, the device provides an immersive experience. Participants simply had to set up the room setup once, enabling them to get started right away. The only extra step was to adjust the head strap for comfort. 
Both games featured Hand Tracking 2.0, the most advanced hand tracking technology available from Meta at the time of the research. This system performed well in reliably tracking hand motions, especially in difficult situations like fast gestures or short hand-to-hand contact. 

The research used a within-subjects design, which ensured that every participant encountered each condition. 
The combination of differed gaming durations, including both short and long sessions, and two control modalities (controllers and hand tracking), resulted in four separate experimental conditions with the order of conditions set by a Latin square layout.
The study's procedure included a number of brief introduction sessions. Whenever participants began a new gaming session, they were given a short description of the setting and the tasks they were given, supported by illustrations such as screenshots. Similarly, control method adjustments were accompanied by short instructions that included controller use demos, button functionality, and explanations of hand tracking movements.
Following each gaming session, participants were asked to answer a series of UX questions, finishing in a final post-questionnaire at the end of the research. 

%Questionnaires 

The demographics part of pre-questionnaire asked participants about their gender, age, employment, and self-assessment of their VR experience. Responses were scored on a scale of 1 ("not at all") to 5 ("very experienced"). %
The part of pre-questionnaire was the standardised ATI (Affinity for Technology Interaction) questionnaire. This questionnaire had nine questions, each with a 6-point Likert scale ranging from 1 ("completely disagree") to 6 ("completely agree"). The ATI score was calculated by taking the average of all item scores and adjusting for three items that were reverse-worded in compared to the others \cite{franke2019personal}.

Following, several UX questionnaires were used to measure different ascepts of UX for each condition. 

\begin{itemize}
    \item The igroup presence questionnaire (IPQ) was used to assess presence, with questions scored on a 7-point Likert scale and altering anchor points \cite{schubert2001experience}.
    \item In addition, the Self-Assessment Manikin (SAM) was used to assess different emotional responses. SAM used a 5-point Likert scale and had a nonverbal, picture-oriented design, that included several sorted variants of an image reflecting different aspects. This questionnaire, well-established and notably concise, has found application across diverse contexts, offering an advantage when presented alongside a variety of questionnaires.
    \item The final UX questionnaire used was the Short User Experience Questionnaire (UEQ-S), which is a simplified version of the User Experience Questionnaire (UEQ). This condensed version reduced the original 26-item collection to only 8, while additionally lowering the number of categories from six to two: pragmatic and hedonic quality. The average score from both of these categories might be viewed as an overall assessment \cite{laugwitz2008construction}.
    \item Lastly, the Flow State Scale (FSS) was used to assess the state of flow, representing a comprehensive questionnaire with 36 individual items distributed across the 9 dimensions of flow \cite{jackson1996development}.

\end{itemize}

In a subsequent post-questionnaire, participants were given the opportunity to express their willingness to recommend VR controllers and hand tracking to others. They could also provide reasoning for their decisions and offer feedback on the overall study experience.

\subsection{Participants}

The study included a total of 20 participants, recruited via the institution's online portal for test subjects. Within the sample of 20 individuals, 7 identified as female and 13 as male. In particular, 65\% of the participants identified as students, which matches our predictions given the recruiting methods. The participants' average age was 28.65 years, with the youngest being 20 years old and the oldest being 57 years old, for a standard deviation of 9.25. Additionally, the mean level of VR experience, measured on a scale of 1 ("not at all") to 5 ("very experienced"), was 2.50, with a standard deviation of 0.95.

\section{Results}
To analyze the collected questionnaire data, two separate statistical approaches were used.
The results from the UX questionnaires (IPQ, SAM, UEQ-S, FSS) were analysed using a two-way repeated measures analysis of variance (ANOVA). This statistical test determines if two variables have a statistically significant interaction impact on a continuous dependent variable. To reduce the probability of type I errors, Bonferroni correction was used. 
The ANOVA's assumption of normality was tested using the Shapiro-Wilk test, which is considered a more trustworthy technique than utilising raw data \cite{kozak2018s}. It is worth noting that non-normal data was obtained; yet, ANOVA is often recognised as resilient in the face of departures from the normality assumption. The research results show that there are no significant effects on type I error rates \cite{blanca2017non}, confirming the validity of using ANOVA even in the absence of strict normality \cite{norman2010likert}.
For the remaining dataset, including the ATI score, the VR experience rating, and the hand tracking recommendation, a binomial logistic regression analysis was performed.

\subsection{Presence}
\begin{figure*}[h!]
  \centering
   \subfloat[IPQ and Gameplay Duration]{
     \includegraphics[scale=0.22]{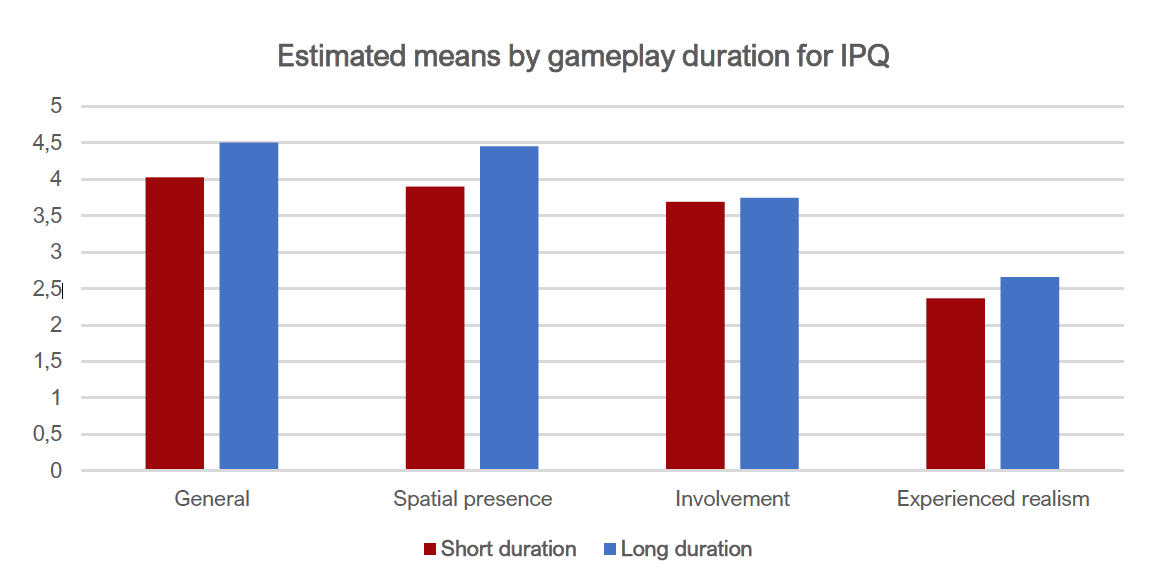}
    }
  \subfloat[UEQ-S and Control Methods]{
    \includegraphics[scale=0.22]{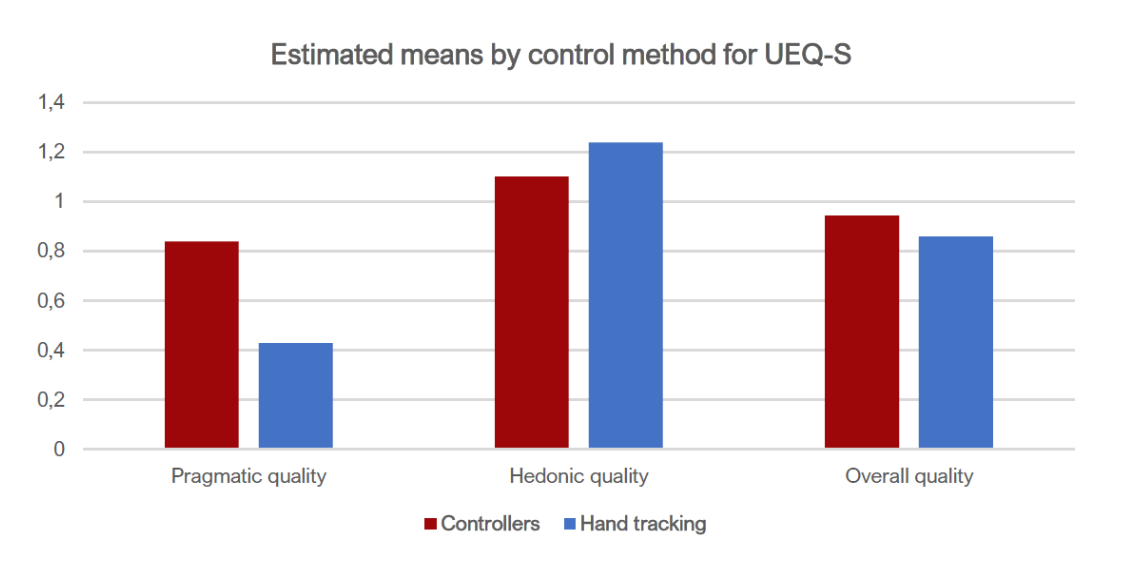}
    }
    \caption{Overview of results: a) Chart depicting estimated marginal means by gameplay duration for IPQ dimensions, b) Chart depicting estimated means by control method for UEQ-S dimensions.}
    \label{fig: results}
\end{figure*}

Significant differences were revealed across all dimensions of the IPQ questionnaire when comparing different gameplay duration conditions. Before proceeding with the analysis, the original 1/7 scale used in the study was converted to a 0/6. Figure \ref{fig: results} provides a complete overview of how gaming duration affects the IPQ dimensions.

In terms of general presence, gaming time had an effect, with a significant difference between short and long durations (F(1,19) = 7.006, p =.016, partial $\eta_{}^{2}$ =.269). Overall, the long duration condition (4.500±0.185) was reported to make users feeling more in presence compared to the short duration condition (4.025±0.225), with a mean difference of 0.475 (95\% CI, 0.099 to 0.851).
In the context of spatial presence, a statistically significant main effect of gameplay duration was found as well (F(1,19) = 19.413, p = .001, partial $\eta_{}^{2}$ = .505). Spatial presence increased significantly in the long duration condition (4.450±0.142) compared to the short duration condition (3.895±0.184), with a mean difference of 0.555 (95\% CI, 0.291–0.819).
Involvement was significantly higher when using controllers (3.925±0.249) than hand tracking (3.450±0.300) for short durations (F(1,19) = 7.228, p =.015), with a mean difference of 0.475 (95\% CI, 0.105 to 0.845). However, long-term participation with controllers (3.738±0.280) did not show a statistically significant difference from hand tracking (3.750±0.282) (F(1,19) = 0.005, p =.944). 
In the scope of experienced realism, the main effect of gameplay duration produced statistical significance (F(1,19) = 4.367, p = .050, partial $\eta_{}^{2}$ = .187). Experienced realism increased significantly in the long length condition (2.656±0.237) compared to the short duration condition (2.363±0.237) where users due to less time to play have felt environment is less realistic, with a mean difference of 0.294 (95\% CI, 0 to 0.588).

\subsection{Pragmatic Quality}
%Pragmatic Quality
Pragmatic quality, one of the dimensions of the UEQ, showed significant differences in this research, resulting in it being the only dimension within the UEQ to show statistical significance. Pragmatic quality evaluates a system or interface's usefulness, efficiency, and usability from the standpoint of the user. 
The study's 1/7 scale was transformed into a $-3$/$+3$ scale prior to analysis. 

The main effect of control method showed a statistically significant difference between controllers and hand tracking (F(1,19) = 6.252, p =.022, partial $\eta_{}^{2}$ =.248). The study found that controllers (0.938±0.156) outperformed hand tracking (0.569±0.130) in terms of pragmatic quality, with a mean difference of 0.369 (95\% CI, 0.060 to 0.677). Figure \ref{fig: results} shows an overview of how control methods effected the UEQ-S dimensions.

\subsection{Clear Goals, Concentration and Sense of Control}
The study of the Flow State Scale (FSS) data revealed significant results in important elements of the flow experience. Particularly, participants' assessments for specific goals, attention, and sense of control changed significantly throughout the analysed activities, giving insight on these important features of the flow state.

The study revealed a statistically significant difference in the main impact of gameplay length (F(1,19) = 7.404, p =.014, partial $\eta_{}^{2}$ =.280) on the dimension associated with specific goals. Clear goals were significantly greater during short length sessions (4.488±0.109) than long duration sessions (4.019±0.163), with a mean difference of 0.469 (95\% CI, 0.108 to 0.829).
When it comes to the focus on the task, the main effect of control method was statistically significant (F(1,19) = 5.000, p =.038, partial $\eta_{}^{2}$ =.208), indicating that controllers (4.475±0.109) were significantly more focused on the task at hand than hand tracking (4.350±0.124), with a mean difference of 0.125 (95\% CI, 0.008 to 0.242).
The main effect of gaming time was statistically significant (F(1,19) = 6.491, p =.020, partial $\eta_{}^{2}$ =.255) in terms of the feeling of control dimension. Short length sessions resulted in a stronger sense of control (4.238±0.102) compared to long duration sessions (3.813±0.196), with a significant mean difference of 0.425 (95\% CI, 0.076-0.774). 
The main effect of control method was statistically significant (F(1,19) = 15.073, p =.001, partial $\eta_{}^{2}$ =.442). Additionally, controllers were reported to have a significantly higher sense of control (4.281±0.121) than hand tracking (3.769±0.170), with a mean difference of 0.513 (95\% CI, 0.236 to 0.789).

\subsection{Previous VR experience}
In terms of participants' past VR experience, a binomial logistic regression was used to determine its impact on the probability of recommending hand tracking technology. In this respect, the model explained 26.1\% of the variation in hand tracking suggestions while correctly classifying 74.1\% of instances.
It is worth mentioning that the predictor variable, VR experience, was statistically significant (p =.042). This finding points out to a relationship: as individuals' levels of VR experience increased, their tendency to recommend hand tracking to others showed a significant rise.

\section{Discussion}
The study's findings provide intriguing insights into how gaming time and control approaches affect VR user experience. While controllers received recognition for their precision and ease of use, hand tracking received mixed reviews, with some appreciating its inventive potential but others criticising its current technological limits.

\subsection{Feedback on control methods}

Participants provided their opinions on the control methods and hand tracking through a concluding survey, after testing each for approximately 12 minutes. The controllers received equally positive feedback, including recognition for their precision and reliability. Participants rated it easier to use, with many noticing that grabbing items in VR felt very comparable to real-world interactions. In contrast, evaluations on hand tracking differed. It was recognised as an innovative and exciting technology that provided a better level of immersion and realism by allowing users to see their hands in the virtual world. However, it was stated that the technique required additional refinement. There were issues with its accuracy and the strange feeling caused by delay. The need to keep hands inside the camera's view was criticised, and the lack of actual sensation when grabbing digital objects. The tactile sense provided by controller buttons was preferred to the absence of feedback in hand tracking. Observations throughout the study revealed challenges with hand-to-hand interactions and the cameras' narrow field of vision, resulting in unpredictable motions when hands went out and then back into the monitored region. Hand tracking also did not work with precision activities like turning items, which significantly impacted the gameplay experience.

\subsection{User Experience Insights}

The data mainly showed the independent effects of gaming duration and control method on several metrics, with significant effects detected. Longer gaming durations increased measures of presence and realism, indicating that prolonged VR experiences improve the perception of being in a real environment. Surprisingly, despite user feedback, hand tracking had no equivalent effect on perceived realism. The study noted differences in clarity of objectives and control sensation between short and long gameplay sessions, potentially influenced by the nature of the games used. Controllers were shown to considerably improve task attention, probably due to experience with comparable gaming gadgets and their inherent reliability. The lack of tactile feedback in hand tracking has a negative effect on user experience, highlighting the advantages of controllers for replicating realistic interactions. Gameplay duration also played a significant role in immersion and presence, with longer sessions resulting in better outcomes.

General presence, spatial presence, and experienced realism, all of which were measured using the IPQ, were statistically significantly higher for the long gameplay duration. This could indicate that spending more time in a VR experience helps to convince the player of being a part of a real environment, instead of just playing a game. 
Interestingly, hand tracking did not show a comparable effect on the experienced realism, even though multiple participants explained that the control method felt more real in the post-questionnaire. However, the goals seemed clearer, and the sense of control was statistically significantly improved for the short gameplay duration, as measured using the FSS. This could be ascribed to the nature of the two games. Vacation Simulator is a bit more expansive compared to Cubism, which may have had an effect on the Clear goals dimension. Additionally, Vacation Simulator requires the use of two hands for some scenarios, which perhaps impacted the sense of control due to players typically using just a single hand for Cubism as observed during the study. This is not in line with another study that indicated higher flow for the longer duration. However, that experiment also labelled 2 minutes as the short duration and 5 minutes as the long duration, as opposed to 3 and 9 minutes here \cite{volante2018time}. 

Hedonic quality however was not impacted by gameplay duration or control method, even though controllers were objectively and subjectively worse. Both were measured using the UEQ-S. Sense of control was statistically significantly higher for controllers as well, in addition to the effect caused by the gameplay duration, which probably also stems from the reliability discrepancy and was measured using the FSS. Participants commented that hand tracking does not feel natural at all due to the delay between moving one’s hands and seeing the result in VR. The average temporal delay for hand tracking is a significant 38.0 milliseconds \cite{abdlkarim2022methodological}. Measured using the IPQ, involvement showed a two-way interaction effect and was higher for controllers for the short gameplay duration. This could again have been influenced by the controller’s superior reliability. 

\subsection{VR Experience Level and ATI Score}

The study also aimed to investigate the relationship between participants' VR experience and their willingness to suggest hand tracking. Interestingly, people with greater VR expertise were more likely to recommend hand tracking, contrary to our hypothesis. This might indicate that experienced users are more willing to accept the limits of existing VR technology. Participants with the least experience reported regular problems with hand tracking, but those with the most experience did not, indicating a better tolerance or acceptance. The technology interaction affinity (ATI) score had no significant effect on hand tracking suggestions, indicating that other characteristics were not evaluated. 

\subsection{User Study Limitations}

A critical limitation of the study was the use of different games for varying gameplay durations, driven by the impracticality of developing a custom VR application. The games were chosen based on compatibility with both control techniques, ethical acceptability, and beginning accessibility, resulting in the choice of using different games for short and long sessions. This provided a variable that might influence the perception of findings, particularly regarding feeling of presence and realism due to different game design. The short-duration game Cubism and the long-duration game Vacation Simulator presented distinct experiences that might impact user feedback and performance measures, needing additional caution when using these outcomes.

\section{Conclusion}

In conclusion, our findings align with the expected results regarding the comparison between controllers and hand tracking within virtual reality (VR) environments. As assumed, the comparison of controllers and hand tracking revealed that controllers are the more rational choice in every way, at least for the specified VR games, corroborating previous findings \cite{hameed2021evaluating}. This advantage is attributed to the controllers' more reliability, their perception of control, and users' increased task attention, all of which lead to a more immersive VR experience. However, it is important to note that previous study has shown that hand tracking can outperform controllers for some interactions \cite{voigt2020influence}, demonstrating a context-dependent preference that changes with the nature of the interaction and the users' experience with VR technology. The mixed findings point to a dynamic environment for VR interaction approaches, with the option between controllers and hand tracking potentially evolving as technology progresses and user experiences expand. Future research is encouraged to further explore these interactions and the potential shifts in user preference as the fidelity of hand tracking improves.

\bibliographystyle{ACM-Reference-Format}
\bibliography{references}

\end{document}